\documentclass[11pt,english]{article}
\pdfoutput=1
\usepackage{jheppub}
\usepackage{amsfonts}
\usepackage{bbm}
\usepackage{multirow}
\usepackage{babel}
\usepackage{verbatim}
\usepackage{mathrsfs}

\usepackage{amsmath}
\usepackage{verbatim}
\usepackage{graphicx}
\usepackage{slashed}
\newcommand{\mg}{\mathcal{G}}

\newcommand{\be}[1]{ \begin{equation}\label{#1} }
\newcommand{\ee}{\end{equation}}
\newcommand{\bea}{\begin{eqnarray}}
\newcommand{\eea}{\end{eqnarray}}
\newcommand{\besu}{\begin{subequations}}
\newcommand{\eesu}{\end{subequations}}
\newcommand{\bes}{\be{}\begin{split}}
\newcommand{\ees}{\end{split}}

\newcommand{\p}{\partial}
\renewcommand{\a}{\alpha}
\renewcommand{\b}{\beta}

\newcommand{\non}{\nonumber}

\newcommand{\refb}[1]{(\ref{#1})}

\renewcommand{\>}{\rangle}

\def\beaa{\begin{eqnarray*}}
\def\eeaa{\end{eqnarray*}}
\def\bee{\begin{equation*}}
\def\eee{\end{equation*}}
\def\bea{\begin{eqnarray}}
\def\eea{\end{eqnarray}}
\def\be{\begin{equation}}
\def\ee{\end{equation}}
\def\ba{\begin{align}}
\def\ea{\end{align}}

\title{Flatspace Chiral Supergravity}

\author[a]{Arjun Bagchi,} \author[b, c, d]{Rudranil Basu,} \author[b] {St\'{e}phane Detournay,} \author[a]{ and Pulastya Parekh.} 

\author{\\}

\affiliation[a]{Indian Institute of Technology Kanpur, Kalyanpur, Kanpur 208016. INDIA. \\}

\affiliation[b]{Universit\'{e} Libre de Bruxelles and International Solvay Institutes, Campus Plaine C.P. 231, B-1050 Bruxelles, Belgium \\} 

\affiliation[c]{Saha Institute of Nuclear Physics (HBNI), Block-AF, Sector-1, Salt Lake, Kolkata 700064, India. \\}
\affiliation[d]{Theoretische Natuurkunde, Vrije Universiteit Brussel, Pleinlaan 2, B-1050 Brussels, Belgium.\\} 

\emailAdd{abagchi@iitk.ac.in, rudranil.basu@ulb.ac.be, sdetourn@ulb.ac.be, pulastya@iitk.ac.in}

\abstract{We propose a holographic duality between a 2 dimensional (2d) chiral superconformal field theory and a certain theory of supergravity in 3d with flatspace boundary conditions that is obtained as a double scaling limit of a parity breaking theory of supergravity. We show how the asymptotic symmetries of the bulk theory reduce from the ``despotic" Super Bondi-Metzner-Sachs algebra (or equivalently the Inhomogeneous Super Galilean Conformal Algebra) to a single copy of the Super-Virasoro algebra in this limit and also reproduce the same reduction from a study of null vectors in the putative 2d dual field theory. }

\preprint{}

\begin{document}

\maketitle

\section{Introduction}
The Holographic Principle offers us a path to a quantum theory of gravity through a non-gravitational field theory in one lower dimension. The AdS/CFT correspondence \cite{Maldacena:1997re} is its best understood avatar, but it is believed that holography is more general and should hold for all spacetimes. Over the last several years, the original Maldacena proposal has been extended away from its familiar relativistic setting in AdS to include non-relativistic holography \cite{Son:2008ye, Balasubramanian:2008dm, Kachru:2008yh}, higher spin holography \cite{Klebanov:2002ja, Sezgin:2002rt, Gaberdiel:2010pz} and gauge-gravity dualities in de Sitter spacetimes \cite{Witten:2001kn, Strominger:2001pn}. 

Using the notion of asymptotic symmetries at the null boundary of spacetime characterized by the Bondi-Metzner-Sachs (BMS) group \cite{Bondi:1962px, Sachs:1962zza, Barnich:2006av}, holography for asymptotically flat spacetimes \cite{Barnich:2010eb, Bagchi:2010zz} has recently met with a certain number of successes, 
though the discussion has often been confined to three dimensions and with theories without supersymmetry. An incomplete list of important works in this direction is here \cite{Bagchi:2012cy} -- \cite{Barnich:2015uva}. We point the reader to \cite{Bagchi:2016bcd, Riegler:2016hah} for a summary of the state of the field. Supersymmetry is crucial to the original correspondence in AdS and is a feature we wish to retain as we build towards a string theoretic understanding of flat holography. Some recent efforts at supersymmetrization of the results in 3d include \cite{Barnich:2014cwa, Barnich:2015sca, Lodato:2016alv, Banerjee:2017gzj}. 

In this paper, we propose a holographic duality between a specific supersymmetric gravitational theory in 3d and a 2d chiral superconformal field theory. The theory of gravity is a supersymmetric extension of 3d Einstein gravity in the first order Chern-Simons formulation (an analogous theory with minimal supersymmetry was named `reloaded' elsewhere, e.g. in \cite{Barnich:2014cwa}), with asymptotically flat boundary conditions, which we obtain as a double scaling limit of a parity breaking theory of supergravity, which is analogous to a supersymmetric version of Topologically Massive Gravity. The double scaling limit requires us to send the coefficient $\mu$ of the gravitational Chern-Simons term to zero and tuning Newton's constant $G \to \infty$, while keeping the product $\mu G$ finite. 

On the field theory side, we work with the assumption that the dual theory inherits the asymptotic symmetry algebra as its underlying symmetry. For our parity breaking theory, like the case of usual supergravity with asymptotically flat boundary conditions, this turns out to be extended ($\mathcal{N}=2$) versions of the Super-BMS$_3$ algebra or equivalently the 2d Super Galilean Conformal Algebra (SGCA). 

We note here, following \cite{Lodato:2016alv}, that there are two distinct supergravity theories that one can obtain in 3d flatspace. The asymptotic symmetries for both these theories are supersymmetric extensions of the BMS$_3$ algebra (or the GCA$_2$) which we discuss below in Eq \refb{bms}. As per the nomenclature of \cite{Lodato:2016alv}, we will be interested in the twisted or ``despotic" supergravity (the asymptotic symmetry algebra of is the Inhomogeneous SGCA \refb{sgcai}) as opposed to the usual Poincare supergravity in 3d, also rather wonderfully named ``democratic" in \cite{Lodato:2016alv} (for which the Homogeneous SGCA \refb{sgcah} appears as asymptotic symmetries). 

The initial departure of our analysis from the case of \cite{Lodato:2016alv} is that for the parity breaking ``despotic" theory, the asymptotic symmetry algebra has two non-zero central extensions in our case, instead of a single one. We perform a scaling limit on the field theory with Super-BMS symmetries and, through an analysis of null vectors, show that there is a consistent truncation to a single copy of a Super-Virasoro algebra. The putative field theory dual is hence governed by the symmetries of a Super-Virasoro algebra and is thus a 2d chiral Super-Conformal field theory (SCFT). The calculation of charges on the bulk side yields results consistent with this, with the charges that correspond to the other generators of the Super-BMS$_3$ algebra identically vanishing in the scaling limit from the initial parity breaking theory.      

We thus propose a duality between a chiral theory of  3d Supergravity that is a Supersymmetric extension of Chern-Simons gravity, with flat boundary conditions, which we will call Flat Space Chiral Supergravity, and a 2d chiral SCFT with a certain central charge. 

Our present analysis can be looked upon as a supersymmetric generalisation of the Flatspace Chiral Gravity story \cite{Bagchi:2012yk}. The asymptotic symmetries of 3d flat space at null infinity is the BMS$_3$ algebra:
\begin{subequations}\label{bms}
\bea
&& [L_m, L_n] = (m-n)L_{m+n} + \frac{c_L}{12} m^3 \delta_{m+n,0} \\
&& [L_m, M_n] = (m-n)M_{m+n} + \frac{c_M}{12} m^3 \delta_{m+n,0} \\
&& [M_n, M_m] = 0.
\eea
\end{subequations}
Here $M_n$'s are translations of the null direction which depend on the angle at the boundary and are called supertranslations. $L_n$'s are the diffeomorphisms of the circle at the boundary and are called superrotations. In Einstein gravity, the central charges take the values $c_L=0$ and $c_M= 3/G$ \cite{Barnich:2006av}. These symmetries can be looked upon as the symmetries of a putative dual 2d field theory living on the null boundary of flatspace.  Although these symmetries have shown up in various contexts, like the non-relativistic limit of AdS \cite{Bagchi:2009my}, Galilean gauge theories \cite{Bagchi:2014ysa, Bagchi:2015qcw, Bagchi:2017yvj}, and also in the tensionless limit of string theory \cite{Bagchi:2013bga, Bagchi:2015nca} and relatedly in ambitwistor strings \cite{Casali:2016atr}, concrete examples are difficult to come by, with the notable exception of \cite{Barnich:2012rz}. It would be much easier to have examples if the symmetry algebra was simply the Virasoro algebra. This requires one to find a truncation of \refb{bms} down to its Virasoro sub-algebra. This is achieved by first looking at a bulk theory which is TMG instead of Einstein gravity. The asymptotic symmetries of TMG with flatspace boundary conditions again yields the BMS$_3$ algebra \refb{bms}, but now the central terms both become non-zero. It is then possible to perform a double scaling on the theory so that we can get $c_L \neq 0$ but $c_M=0$. On the bulk side, this reduces TMG to Chern-Simons gravity. On the boundary, through an analysis of null vectors, it is possible to show that this limit enables one to achieve the desired truncation down to a single copy of the Virasoro algebra. It was thus conjectured that Chern-Simons gravity with flat space boundary conditions is dual to the chiral half of a 2d CFT. More specifically, connections were made to a specific dual theory, a monster CFT. For more details, the reader is referred to \cite{Bagchi:2012yk}. 

In our present paper, we attempt a supersymmetric version of the analysis reviewed above. Here is a brief outline of the rest of the paper. In Sec 2, we present our bulk theory which is a parity violating theory of supergravity. We calculate the charges and the asymptotic symmetry algebra for this theory and then perform our scaling limit. The next section is devoted to the analysis on the field theory side, which is a detailed discussion of the null vectors in a 2d theory invariant under Super BMS$_3$. We show that under the proposed scaling, the representations of the field theory side reduce from modules of super-BMS to super-Verma modules of a single copy of a Super Virasoro algebra. We end with a summary of our results and some discussions. There are three appendices supplementing the calculations performed on the bulk theory and one with some details of the boundary theory. 

\section{Parity Breaking Supergravity}
\subsection{Dynamics of $\mathcal{N}=2$ supergravity}
The global infinitesimal isometries of 3d flat space form the non-semisimple Lie algebra $\mathfrak{g}=\mathfrak{iso}(2,1)$ generated by 3 homogeneous Lorentz generators and translations generators along the 3 space-time directions. In a basis convenient for the present purpose, this reads:
\bea \label{alg}
[J_m, J_n] = (m-n)J_{m+n}, ~~ \, [J_m, P_n] = (m-n)P_{m+n}, ~~ \, [P_m, P_n] = 0
\eea
for $m,n = -1,0,1$. It is an age-old idea that this algebra could be gauged to find a theory of gravity with a vanishing cosmological constant. Building towards this goal, one defines a $\mathfrak{g}$ valued connection 1-form and the Chern-Simons form constructed out of it is expected to give a theory of gravitation. Taking this a bit further, one can inquire whether supergravity theories with built-in local supersymmetric invariance can be obtained by similar arguments. The answer, as expected, is affirmative. The key here lies in finding a Lie-superalgebra whose bosonic part is $\mathfrak{g}$, i.e. \eqref{alg}.
 
To this end, let us introduce a couple of fermionic generators $S_{\alpha}, R_{\alpha}$. 
The $\alpha = \pm \frac{1}{2}$ index reflect the 2d representation of the 3d Clifford algebra. The non-vanishing brackets are (apart from those spelled out in \eqref{alg}):
\bea \label{alg2}
&& [J_m, S_{\alpha}] = (m/2- \alpha) S_{m+ \alpha}, ~~\, [J_m, R_{\alpha}] = (m/2- \alpha) R_{m+ \alpha} \non \\
&& [P_m, S_{\alpha}] = (m/2- \alpha) R_{m+ \alpha}  \\
&& \{S_{\alpha}, S_{\beta}\} = J_{\alpha+ \beta}, ~~ \, \{S_{\alpha}, R_{\beta}\} = P_{\alpha+ \beta} \non
\eea
Let us call this superalgebra $\tilde{\mathfrak{g}}$. The algebra valued connection 1-form can thus be expressed in this basis as:
\bea \label{connection}
A = e^n P_n + \omega^n J_n + \frac{1}{\sqrt{2}}\left(\psi^{\alpha} S_{-\alpha} + \eta^{\alpha} R_{-\alpha}\right) .
\eea
Here $e, \omega$ respectively are the vielbein and spin connection 1-forms, while the fermionic fields $\psi$ and $\eta$ stand for the Majorana gravitino fields. The last input required to construct (an action of) a gravitational theory is an invariant quadratic form on the algebra. It is well known that the bosonic part $\mathfrak{g}$ being non-semisimple, the canonical choice of Killing metric is degenerate. However that problem can easily be avoided by defining \cite{Achucarro:1987vz, Witten:1988hc}:
\bea \label{inn_prod}
\langle J_m, P_n \rangle = \gamma_{mn}, \ \ \mbox{ where } \gamma = \mathrm{anti}\mbox{-}\mathrm{diagonal} (-2, 1 , -2)
\eea
Moreover it is augmented for the whole $\tilde{\mathfrak{g}}$ by the portion involving the super-trace of the fermionic generators:
\bea \label{charg_conju}
&& \langle S_{\alpha}, R_{\beta} \rangle =  C_{\alpha \beta}, \ \ \mbox{with} \ \ C = \left( {\begin{array}{cc}
   0 & -2 \\
   2 & 0 \\
  \end{array} } \right). 
\eea 
Here $C$ is the charge conjugation matrix. With respect to the above inner product, the Chern-Simons action:
\bea \label{csaction}
S= \frac{k}{4 \pi}\int \langle A \wedge dA + \frac{2}{3} A\wedge A \wedge A \rangle
\eea
($k = \frac{1}{4G_N}$ is the Chern-Simons level with $G_N$ being the 3d Newton's constant) now in terms of the gravitational fields takes the standard $\mathcal{N}=2$ supergravity form:
\bea \label{gravaction}
S= \frac{1}{16 \pi G_N}\int 2 e^n \wedge R_n - \bar{\psi} D \eta - \bar{\eta} D \psi + \frac{1}{2} \bar{\psi} e^n \Gamma_n \psi .
\eea
Here $D$ as usual is the covariant derivative with respect to the connection $\omega$ and we have used the super-Lie algebra $\tilde{\mathfrak{g}}$ and the invariant bilinear form on it \eqref{inn_prod}, \eqref{charg_conju}. As per the present convention, the conjugation of Majorana spinors in \eqref{gravaction} has been defined via the charge conjugation matrix \eqref{charg_conju} and the Gamma matrices are given in terms of the Pauli matrices:
\bea
\Gamma _0 = \sigma_3, \, \Gamma_{\pm 1}= -i\sigma_2 \pm \sigma_1 .
\eea
According to the terminology of \cite{Lodato:2016alv}, \eqref{gravaction} is the ``despotic'' form of flat space supergravity. 

Supersymmetry is built into the action \eqref{csaction} as it is locally gauge invariant (small gauge transformations) and the connection \eqref{connection} is super-Lie algebra valued.  
Equation of motion in the Chern-Simons version is flatness of the connection, which, in terms of the curvature $R_n$ of the spin-connection and the torsion $T^n$, translates to:
\bea
R^n = - \frac{1}{4} \bar{\psi} \Gamma ^n \psi, ~~ \, T^n= - \frac{1}{4} \bar{\psi} \Gamma ^n \eta.
\eea
For the matter fields, the equations of motion are:
\bea
D \psi =0 , ~~ D \eta = - \frac{1}{2} e^n \Gamma_n \psi.
\eea

A striking feature of 3d gravity in the first order formulation is the existence of a one-parameter family of actions, all of which give the same equations of motion. It was most prominently described in \cite{Witten:1988hc} for AdS gravity and later in the context of supergravity in \cite{Giacomini:2006dr}. In the case of interest, {\it{viz.}} asymptotically flat supergravity with various amount of supersymmetry, this has been addressed in in \cite{Barnich:2014cwa, Barnich:2015sca, Basu:2017aqn, Fuentealba:2017fck}. In light of the present analysis, the action \eqref{gravaction} is a single member of the above mentioned family. 

In the case of supergravity, the above can  be extended to a one-parameter family of theories by a simple twist in the Lie algebra inner product. More explicitly this is done by supplementing \eqref{inn_prod}, \eqref{charg_conju} with
\bea \label{mod_inn_prod}
\langle J_m, J_n \rangle  = \frac{1}{\mu} \gamma_{mn}, \, \, \langle S_\alpha , S_\beta \rangle  = \frac{1}{\mu} C_{\alpha \beta}
\eea
for a real parameter $\mu$. It is straightforward to see that with respect to this modified inner product the Chern-Simons action gives rise to an action which contains $\mu$ dependent terms in addition to the earlier action \eqref{gravaction}
\bea \label{mod_action}
\tilde{S}= \frac{1}{16 \pi G_N}\int 2 e^n \wedge R_n + \frac{1}{\mu}\mathrm{CS}(\omega) - \bar{\psi} D \eta - \bar{\eta} D \psi + \frac{1}{2} \bar{\psi} e^n \Gamma_n \psi - \frac{1}{\mu}\bar{\psi} D \psi.
\eea
Here $ CS(\omega) = \langle \omega \, d \omega + \frac{2}{3} \omega ^3 \rangle$ is the Chern-Simons 3-form for the connection $\omega$. This $\mathcal{N}=2$ theory can be viewed in contrast to the parity breaking $\mathcal{N}=1$ supergravity action used in \cite{Barnich:2014cwa}, termed `reloaded' by the authors.

However equations of motion don't alter (with respect to the Chern-Simons theory, they still stem from the flatness of connection based on the same Lie algebra). As long as classical solutions are concerned, therefore, all the members in this family of theories are the same. On the other hand, the charge algebra of large diffeomorphisms is affected because of the modification of the canonical structure. In addition, due to the emergence of a Lorentz-Chern-Simons term, parity is broken.

To illustrate the point regarding the canonical structure, we first remind ourselves of a couple of well known facts about Chern-Simons theory:
\begin{itemize}
\item There are no local physical degrees of freedom. There may however be global degrees of freedom either due to non-trivial topology of background manifold or due to the boundary, if the manifold has one.
\item There is no global rigid symmetry, and hence no conserved quantities associated with them. However gauge invariance may result into non-trivial conserved charges, with support only at the boundary, provided appropriate boundary conditions are met.   
\end{itemize}
The last point can be analytically expressed as a generically non-integrable variation of a charge corresponding to a gauge generator $\Lambda$, a  $\tilde{\mathfrak{g}}$ valued space-time functions:
\bea \label{unint}
\slashed{\delta}Q[\Lambda] = -\frac{k}{2 \pi}\int_{\partial \Sigma} \langle \Lambda, \delta A \rangle.
\eea
Here $\Sigma$ is a 2-surface, which can be treated as a spatial one when a space-time interpretation is attached to the background manifold. This is now clear that both the dependence of the gauge parameter $\Lambda$ on $A$ and the asymptotic data on $A$ determine the existence of a charge $Q$ above \eqref{unint}. It is also to be noted here, if a charge exists, it would, in our specific theory, depend on the twist or parity breaking parameter $\mu$ via the inner product.

From the gravity perspective, there are only diffeomorphism invariance of the theory generated by space-time vector fields. In case of supergravity, like the one we are interested in, there are local supersymmetric invariance as well. For a diffeomorphism and local supersymmetry transformation generated by $\Xi^{\mu}$ there is an equivalent (on-shell) gauge transformation in the CS picture:
\bea \label{diff-gauge}
\Lambda = \Xi^{\mu} A_{\mu}\eea
which is linearly dependent on the field configuration. In the Appendix \ref{secpre-symp}, we have adapted a covariant phase space analysis of CS theory. There we describe in detail the obstruction to integrability of charges and its resolution for linearly dependent gauge parameters by choosing a mild gauge fixing condition on the asymptotic gauge field.

\subsection{Asymptotic symmetries of $\mathcal{N} =2$ supergravity: Boundary conditions}
We have just observed that all the interesting features in a topological theory like 3D gravity or equivalently CS emerges from the asymptotic boundary. In particular, determining the physical charges, if they exist at all, \eqref{unint} requires specifying field configurations near the boundary.

As is understandable, various physical situations impose strict restrictions on boundary conditions. In the present asymptotically flat supergravity set up, we would consider one such scenario. 
To be more concrete, let us consider the connection $A_{\mathrm{fs}}$ corresponding to 3d Minkowski space devoid of fermionic degrees of freedom. Then we impose the following boundary condition on generic CS flat-connections:
\bea \label{flat_cond}
\left( A -A_{\mathrm{fs}}\right)\big{|}_{r =\mathrm{constant } \rightarrow \infty} = \mathcal{O}(r^0)
\eea
For the gravity interpretation to be clear, we specify the topology of the 2-dimensional null boundary of the background space-time manifold to be a cylinder and coordinatize it by the retarded time $u$ and periodic coordinate $\phi$. The spatial foliations $\Sigma$ with coordinates $(r, \phi)$ are chosen to be discs which cut the null infinity $(r \rightarrow \infty)$ at constant $u$. 

It is evident that the elements of the reduced phase space $\mathcal{P}_{\mathrm{red}}$ defined via \eqref{A-red} satisfy the condition \eqref{flat_cond}. Therefore we can write these connections as:
\bea
A= b^{-1} \left( d + a\right)b.
\eea
For the present work, we further reduce the space of connections. This reduction corresponds to asymptotically flat space-times in `BMS-gauge' \cite{Barnich:2012aw} adapted to include $\mathcal{N}=2$ supersymmetry:
\begin{subequations} \label{bc}
\bea 
b &=& e^{\frac{r}{2}P_{-1}} \\
a &=& \left(P_{+1} - \frac{\mathcal{M}}{4} P_{-1} +\frac{\psi}{4} R_{-1/2}\right)du \non \\
&&+ \left( J_{+1} - \frac{\mathcal{M}}{4} J_{-1} -\frac{\mathcal{N}}{4} P_{-1} + \frac{\psi}{4} S_{-1/2} + \frac{\eta}{4} R_{-1/2}\right) d \phi \label{a}
\eea
\end{subequations}
where $\mathcal{M}, \mathcal{N}$ are bosonic and $\psi, \eta$ (not to be confused with the two component gravitinos appearing in \eqref{connection} or \eqref{gravaction}) are fermionic variables, supposed to coordinatize the phase-space. Hence,
\bea
A &=& \frac{1}{2} P_{-1} dr + (r P_0)d \phi + a
\eea
Flatness of the connection $A$, i.e. on the space of gravitational solutions, the above fields get more restricted and should take the following form:
\bea
\mathcal{M} = \mathcal{M}(\phi),\, \mathcal{N} = \mathcal{J}(\phi) + u\,\mathcal{M^{\prime}}(\phi),\, \psi = \Psi(\phi),\, \eta = \Theta (\phi) + u\,\Psi^{\prime} (\phi).
\eea
Here the primes denote derivative with respect to $\phi$.

These are essentially the boundary conditions presented in \cite{Lodato:2016alv}. Note that all the Lie-(super) algebra components of $a$, by these boundary conditions, are no longer dynamical (in the sense that they ae phase space constants). In the covariant phase space framework of section \ref{charge_algebra_proof}, this is further reduction of the phase space to $\tilde{\mathcal{P}}_{\mathrm{red}} \subset \mathcal{P}_{\mathrm{red}}$ defined by the conditions like $\delta \langle a_u, J_{-1}  \rangle = 0 = \delta \langle a_u, P_{-1}  \rangle$ etc. 

In the specific gauge \eqref{A-red}, for any linearly state-dependent gauge transformation like \eqref{diff-gauge}, existence of an integrable charge is guaranteed if the term:
\begin{align} \label{integ}
\langle a_u, \delta a_\phi \rangle
\end{align}
is a total variation. The interested reader may consult \eqref{obstruc2} for an explanation. A nice feature of the boundary configuration \eqref{a} indeed satisfies this condition. This directly implies that any diffeomorphism (and local supersymmetry transformation) for boundary field configuration satisfying integrability of \eqref{integ} gives a conserved charge supported at boundary, provided it is finite for $r \rightarrow \infty$.

Since we have kept the asymptotic field configurations (boundary fall-off conditions) \eqref{a} the same as in \cite{Lodato:2016alv}, we can freely use some of the relevant results from there. However our further results will differ from theirs, because the dynamic content of the theory we are using, including the canonical structure differ.
\subsection{Asymptotic symmetries of $\mathcal{N} =2$ supergravity: Dynamical realization}
In the above discussion we have noticed that with our boundary conditions \eqref{a}, the existence of charges corresponding to an arbitrary diffeomorphism is guaranteed provided the charge does not diverge as $r \rightarrow \infty $. We now construct the algebra of charges, induced on the phase space from the algebra of gauge transformations (not be confused with the Lie-algebra of the gauge group). Since there is gauge redundancy, in the canonical formalism this algebra is implemented via a Dirac bracket of charges. Avoiding this route, we present a covariant framework for this in Appendix \ref{charge_algebra_proof}. There we prove that more information is required to ensure closure of an algebra of charges corresponding to arbitrary diffeomorphisms (or rather, the associated gauge parameters). For a generic field configuration, only those charges should be considered whose corresponding gauge generators preserve the integrality criterion of $\langle a_u, \delta a_{\phi} \rangle$. 

The present field configuration  \eqref{a} however is more restrictive and contains components which are phase space constants, e.g. $\langle J_{-1}, a_u\rangle = 0$. These conditions define a subspace of the phase space. Therefore it is natural to only consider gauge transformations that are tangential to this reduced space. For the last example of $\langle J_{-1}, a_u\rangle$, this implies that we should restrict the gauge transformation parameter $\lambda = b \Lambda b^{-1}$ to preserve this, 
 \footnote{Note that, as we have seen, the allowed gauge transformation parameters which preserve $\tilde{\mathcal{P}}_{\mathrm{red}}$ depend linearly on the field components. Hence they can be understood to be composed of a diffeomorphism generating transformation and a pure gauge transformation, as well as local supersymmetry as in \eqref{lambda_def}. However, for the purpose of this article, we will not explicitly use this form to identify the diffeomorphism and local supersymmetry generating vector fields.}:
\begin{align} \label{jda} 
\langle J_{-1}, \delta_{\lambda} a_u\rangle = 0.
\end{align}
In order to make the form of allowed transformations explicit, let us express it in terms of our chosen basis:
\bea
\lambda = \xi^n P_n + \chi^n J_n + \epsilon^{\alpha} S_{\alpha} + \zeta ^{\alpha}R_{\alpha}
\eea
Conditions like \eqref{jda} above put following restrictions on components of $\lambda$:
\besu \label{lambda_components}
\bea 
&& \chi^0 = - \chi^{+1}{'}, \,  ~~~~\chi^{-1} = \frac{1}{2}\chi^{+1}{''} - \frac{\mathcal{M}}{4}\chi^{+1} - \frac{\psi}{8} \epsilon^{1/2},  \\
&& \xi^{0} = -\xi^{+1}{'}, \, ~~~~\xi^{-1} = \frac{1}{2}\xi^{+1}{''} - \frac{\mathcal{M}}{4}\xi^{+1} - \frac{\mathcal{N}}{4}\chi^{+1} - \frac{\psi}{8} \zeta^{1/2} - \frac{\eta}{8} \epsilon^{1/2} \\
&& \epsilon^{-1/2} = - \epsilon^{1/2}{'} + \frac{\psi}{4} \chi^{1}, \,~~~~ 
\zeta ^{-1/2} = - \zeta^{1/2}{'} + \frac{\eta}{4}\chi^{1} + \frac{\psi}{4} \xi^{+1}
\eea
\eesu
Supplemented by these, there are also the following conditions on the functional forms of the 4 independent \footnote{By independence of 2-phase space functions $f$ and $g$, we mean the linear independence of the phase space tangent vectors or the variations $\delta f$ and $\delta g$.} functions :
\bea \label{lambda_components_more}
\chi^{+1} =  Y(\phi), \, \xi^{+1} = T(\phi) + u\,Y'(\phi), \, \epsilon^{1/2} = \epsilon(\phi), \, \zeta^{1/2} = \zeta(\phi) + u\, \epsilon'(\phi)
\eea

We have observed from the boundary conditions \eqref{a}, which ultimately reduce the phase space to two bosonic ($\mathcal{M}, \mathcal{N}$) and two fermionic functions ($\eta, \psi $), that the integrability condition \eqref{obstruc2} for charges is met. The explicit form of the integrated charge corresponding to \eqref{lambda_components}, \eqref{lambda_components_more} is:
\bea \label{int_charge}
Q[\lambda(Y,T, \epsilon, \zeta)] = -\frac{k}{4 \pi } \int_{\partial \Sigma =S^1}\left( (\mathcal{J}+ \frac{1}{\mu} \mathcal{M})Y +\mathcal{M} T + \epsilon (\Theta + \frac{1}{\mu} \Psi)+ \zeta \Psi \right) d \phi
\eea
In order to present the algebra of the Dirac brackets of the charges in conventional form, we would express them as the following modes:
\besu \label{modes}
\bea
L_m &:=& -Q[\lambda(Y=e^{im \phi},0, 0, 0)] = \frac{k}{4 \pi} \int_{S^1} d\phi \left(\mathcal{J}+ \frac{1}{\mu} \mathcal{M} \right) e^{im \phi} \\
M_m &:=& -Q[\lambda(0,T=e^{im \phi}, 0, 0)] = \frac{k}{4 \pi} \int_{S^1} d\phi \,\mathcal{M} \, e^{im \phi} \\
G_r &:=& \sqrt{2} Q[\lambda(0,0,\epsilon=e^{i r \phi},0)] = -\frac{\sqrt{2}k}{4 \pi} \int_{S^1} d\phi \left(\Theta+ \frac{1}{\mu} \Psi \right) e^{ir \phi} \\
H_r &:=& \sqrt{2} Q[\lambda(0,0,0,\zeta=e^{i r \phi})] = -\frac{\sqrt{2} k}{4 \pi} \int_{S^1} d\phi\, \Psi \, e^{ir \phi}
\eea
\eesu
The main goal of the asymptotic analysis is to calculate the Dirac brackets of these Fourier modes. As we have found out the gauge transformations which preserve the reduced phase space, it is now guaranteed that the charge algebra should be closed. The brackets can be easily computed using \eqref{generic_db}:
\bea \label{generic_db}
\{Q[\Lambda_1], Q[\Lambda_2]\} = -\frac{k}{2 \pi} \int_{\partial_{\Sigma}} \left(\langle [\Lambda_1, \Lambda_2] , A\rangle + \langle \Lambda_2 , d \Lambda_1 \rangle \right)
\eea
for generic gauge parameters $\Lambda_{1,2}$ and field configuration $A$.

For example, if we choose $\lambda_1(Y=e^{im \phi},0, 0, 0)$ and $\lambda_2(Y=e^{in \phi},0, 0, 0)$, then the Dirac bracket:
\bea \label{sample}
\{L_m, L_n \} = \Omega \left( \delta_{\lambda_1}, \delta_{\lambda_2} \right)= -i(m-n) L_{m+n}-  i  \frac{k}{\mu} m^3 \delta_{m+n,0}
\eea
The explicit computation of this bracket is presented in the Appendix.
This is easily promoted to quantum commutators by the usual prescription $\{A, B \}_{PB} \to i [A, B] $ (having set $\hbar =1$). 
The full charge algebra (non-zero brackets only) is:
\besu \label{super_bms}
\begin{align} 
[L_m, L_n] &= (m-n)L_{m+n} + \frac{c_L}{12} m^3 \delta_{m+n,0} \\
[L_m, M_n] &= (m-n)M_{m+n} + \frac{c_M}{12} m^3 \delta_{m+n,0} \\
[L_m, G_r] &= \left(\frac{m}{2}-r\right) G_{m+r} \\
[L_m, H_r] &= \left(\frac{m}{2}-r\right) H_{m+r}\\
[M_m, G_r] &= \left(\frac{m}{2}-r\right) H_{m+r} \\
\{ G_r, G_s \} &= 2 L_{r+s} + \frac{c_L}{3} r^2 \delta_{r+s,0} \\
\{ G_r, H_s \} &= 2 M_{r+s} + \frac{c_M}{3} r^2 \delta_{r+s,0}
\end{align}
\eesu
where the central charges are both non-vanishing $c_L = 12 k/\mu = \dfrac{3}{\mu G_N}, c_M = 12 k = \dfrac{3 }{G_N}$. The curly braces for fermion-fermion brackets are the usual anti-commutators.

The infinite bosonic modes of the above algebra form the 3d BMS algebra. The whole set incorporating a couple of fermionic ones is formally same as the `despotic' $\mathcal{N}=2$ super-BMS algebra, as per the nomenclature introduced in \cite{Lodato:2016alv}. The important difference here, of course, is the non-vanishing central charge $c_L$. One should note that even in the presence of more than 1 supersymmetry generator, we don't have a bosonic R-current which may generate a rotation in the $G, H $ plane. This is basically due to the clear asymmetry between the two in the algebra \eqref{super_bms}.

\subsection{Bulk Chiral Limit}
In this subsection we describe a mechanism that leads to a chiral truncation of the algebra \eqref{super_bms} in a manner similar to the bosonic case \cite{Bagchi:2012yk}. The starting gravitational theory or theories are a two parameter family characterized by the Newton's constant $G_N$ and the parity breaking parameter $\mu$. The limit $\mu \rightarrow \infty$ gives us back pure Einstein gravity coupled with two fermions in the  `despotic' manner. The limit works fine at each step of the asymptotic symmetry analysis, definition of the charges and even the level of the charge algebra. As expected, these results in this limit match precisely those of \cite{Lodato:2016alv}. In particular the central charges become:
\bea
\lim_{\mu \rightarrow \infty} (c_L, c_M ) = (0, \frac{3}{G_N})
\eea
It is intriguing to observe however that there exists another limit at the parameter space where the roles of the central charges flip, {\it{i.e.}} $c_L$ is non-zero while $c_M$ vanishes. This sector is probed by the double-scaling limit $\mu \rightarrow 0, \ G_N \rightarrow \infty$ such that the product $\mu G_N = \frac{1}{\kappa} $ is a finite constant. In this limit, the action \eqref{mod_action} written in terms of gravity variables simplifies to:
\bea \label{chiral_action}
\tilde{S}= \frac{\kappa}{16 \pi }\int \mathrm{CS}(\omega) - \bar{\psi} D \psi.
\eea
When applied to the conserved charges \eqref{modes} this limit forces 

\be\label{modes}
L_m =\frac{\kappa}{16 \pi} \int_{S^1} d\phi \, \mathcal{M}  e^{im \phi}, \quad G_r =  -\frac{\sqrt{2}\kappa}{16 \pi } \int_{S^1} d\phi \, \Psi  e^{ir \phi}, \quad M_m  = H_r=0 .
\ee
Here we should remind ourselves that the Chern-Simons level $k$ was equated from the start with $\frac{1}{4G_N}$. Dropping off the two sets of generators now reduces the super-BMS algebra \eqref{super_bms} to a single copy of super-Virasoro algebra with central charge $c_L = 3{\kappa}$.

\section{Chiral Limit: Boundary Side}

A natural prescription for holography for a generic spacetime, drawing inspiration from lessons in AdS/CFT, is to assume that the asymptotic symmetries of the gravity theory would be realized as the underlying symmetry algebra of the putative dual field theory. Following this, the supersymmetric field theory dual to the supergravity theory in the previous section will inherit its asymptotic structure as its defining symmetry. Thus, if there exists a field theory that is holographically dual to the ``despotic" supersymmetric parity violating theory we discussed in the earlier section, it would follow what we call the Inhomogeneous Super Galilean Conformal Algebra (SGCA$_I$):
\bea \label{sgcai}
&& [L_n, L_m] = (n-m) L_{n+m} + \frac{c_L}{12} (n^3 -n) \delta_{n+m,0}, \nonumber\\
&& [L_n, M_m] = (n-m) M_{n+m} + \frac{c_M}{12} (n^3 -n) \delta_{n+m,0}, \\
&& [L_n, G_r] = \Big(\frac{n}{2} -r\Big) G_{n+r}, \ [L_n, H_r] = \Big(\frac{n}{2} -r\Big) H_{n+r}, \ [M_n, G_r] = \Big(\frac{n}{2} -r\Big) H_{n+r}, \nonumber\\
&& \{ G_r, G_s \} = 2 L_{r+s} + \frac{c_L}{3} \Big(r^2 - \frac{1}{4}\Big)   \delta_{r+s,0}, \ \{ G_r, H_s \} = 2 M_{r+s} + \frac{c_M}{3} \Big(r^2 - \frac{1}{4}\Big)   \delta_{r+s,0}.\nonumber
\eea
Note that \eqref{sgcai} is a rewritten version of \eqref{super_bms} with a trivial shift in the generators: $L_n \rightarrow L_n +\frac{c_L}{24} \delta_{n,0}, M_n \rightarrow M_n +\frac{c_M}{24} \delta_{n,0}$. We will have both central terms $c_L$ and $c_M$ turned on. From the bulk theory, 
\be{}
c_L = \frac{3}{\mu G_N}, \quad c_M = \frac{3}{G_N}.
\ee
In the preceding sections, we showed that the gravity theory reduced to a supersymmetric version of Chern-Simons Gravity (CSG) under the double scaling limit 
\be{}
\mu \to 0, \quad G_N \to \infty, \quad \mu G_N = \frac{1}{\kappa},
\ee
where $\kappa$ is a constant. The charges corresponding to $M_n$ and $H_r$ vanished identically in this scaling limit, leading us to believe that for CSG, the dual theory could be governed by just a single copy of the Super-Virasoro algebra. In what follows, through an analysis on null vectors in the field theory with SGCA$_I$ symmetry, we show that the scaling limit that we proposed in the bulk indeed corresponds to a consistent truncation from an Inhomogeneous Galilean Conformal Field Theory (SGCFT$_I$) to a chiral half of a Superconformal field theory in 2d. 

\subsection*{Representation theory}
We are interested in the representation theory of the above algebra \refb{sgcai}. We will label the states by the eigenvalues of $L_0$ and $M_0$: 
\be
L_0|\phi\rangle=\Delta|\phi\rangle, \quad M_0|\phi\rangle=\xi|\phi\rangle. 
\ee
We will work exclusively in the NS sector. Hence the modes of the fermionic generators are half integral and this means that there is no further label on the states of the representation. The algebra determines that $\{L_n, M_n, G_n, H_n \}$ lower the $\Delta$ eigenvalue for $n>0$. We want the spectrum of $\Delta$ to be bounded from below and hence, in close analogy to usual 2d CFTs, we define the notion of a primary state $|\phi\rangle_p$ as one for which $\Delta$ cannot be lowered further: 
\be
L_n|\phi\rangle_p = M_n|\phi\rangle_p= G_n |\phi\rangle_p= H_n|\phi\rangle_p=0.
\ee 
The modules of the algebra would be built out of these primary states by acting with raising operators. 

\subsection*{Null state analysis}
We wish to check the reducibility of the modules that are built out of the primaries, as just described. To do this, we shall examine the possibility of occurrence of null states in the spectrum, $i.e.$ states that are orthogonal to all states in the Hilbert space, including themselves. More details of the analysis of this section can be found in \cite{Bagchi:2017cte} and Appendix \ref{apd}.  

First we list the most general states at the first few levels levels. In the following, the state $|n\>$ represents a state at level $n$. The most general states, upto level 2, are given by
\bea
&& |1/2\rangle = a_1G_{-\frac{1}{2}}|\phi\rangle+a_2H_{-\frac{1}{2}}|\phi\rangle, \\
&& |1\rangle = b_1L_{-1}|\phi\rangle+b_2M_{-1}|\phi\rangle+b_3G_{-\frac{1}{2}}H_{-\frac{1}{2}}|\phi\rangle, \\ 
&& |3/2\rangle = d_1L_{-1}G_{-\frac{1}{2}}|\phi\rangle+d_2L_{-1}H_{-\frac{1}{2}}|\phi\rangle+d_3M_{-1}G_{-\frac{1}{2}}|\phi\rangle +d_4M_{-1}H_{-\frac{1}{2}}|\phi\rangle \nonumber \\
&& \qquad \qquad +d_5G_{-\frac{3}{2}}|\phi\rangle+d_6H_{-\frac{3}{2}}|\phi\rangle, \\
&& |2\rangle = f_1L_{-2}|\phi\rangle+f_2L^2_{-1}|\phi\rangle+f_3L_{-1}M_{-1}|\phi\rangle +f_4L_{-1}G_{-\frac{1}{2}}H_{-\frac{1}{2}}|\phi\rangle +f_5M^2_{-1}|\phi\rangle \nonumber \\
&&\qquad +f_6M_{-2}|\phi\rangle +f_7M_{-1}G_{-\frac{1}{2}}H_{-\frac{1}{2}}|\phi\rangle+f_8G_{-\frac{3}{2}}H_{-\frac{1}{2}}|\phi\rangle+f_9G_{-\frac{1}{2}}H_{-\frac{3}{2}}|\phi\rangle. 
\eea
Now we impose the conditions of these states for being null. It is straightforward to see that a state $|m+r\rangle$, $L_n|m+r\rangle=0$, provided $n>m+r$. Same for $M_n,G_s$ and $H_s$ where $n$ or $s>m+r$.  Below we list the conditions for the states above being null. The non-trivial conditions at each level yield: 
\begin{equation}
\begin{split}
\mbox{Level 1/2:} \quad G_{\frac{1}{2}}|1/2\rangle=2a_1\Delta|\phi\rangle+2a_2\xi|\phi\rangle=0,\quad H_{\frac{1}{2}}|1/2\rangle =2a_1\xi|\phi\rangle=0. \ \ \ \ \ \ \
\end{split}
\end{equation}
\begin{equation}
\begin{split}
\mbox{Level 1:} \quad G_{\frac{1}{2}}|1\rangle&=(b_1-2b_3\xi )G_{-\frac{1}{2}}|\phi\rangle+(b_2+2b_3\Delta)H_{-\frac{1}{2}}|\phi\rangle=0, \\  
H_{\frac{1}{2}}|1\rangle&=(b_1+2b_3\xi)H_{-\frac{1}{2}}|\phi\rangle=0, \quad L_1|1\rangle=2[b_1\Delta+(b_2+b_3)\xi]|\phi\rangle =0, \\  
M_1|1\rangle&=2b_1\xi|\phi\rangle=0. \\
\end{split}
\end{equation}
Similarly, we can find the conditions for null states at higher levels. For levels 3/2 and 2, the details are listed in Appendix \ref{apd}.  We then go on to find the restrictions on the constant coefficients for these states:  

\bigskip

\paragraph{\it{Level 1/2:}}
We have 2 constants $a_1,a_2$ satisfying equations:
\begin{equation}
a_1\Delta+a_2\xi=0, \quad  a_1\xi=0.
\end{equation}
To get a non trivial state we must have $\xi=0$, and the null state is 
\begin{equation}
\begin{split}
|1/2\rangle&=a_2H_{-\frac{1}{2}}|\phi\rangle. \\  
\end{split}
\end{equation}
{\it{Level 1:}} 
We have 3 constants $b_1,b_2,b_3$ satisfying equations:
\begin{equation}
b_2\xi=b_3\xi=0, \quad b_2+2b_3\Delta=0,
\end{equation}
and $b_1=0$. To get a non trivial state we must have $\xi=b_1=0$ and $b_3=-\frac{1}{2\Delta}b_2$. If $\Delta\neq0$, the null state becomes 
\bes\label{w}
|1\rangle&=b_2\Big(M_{-1}|\phi\rangle-\frac{1}{2\Delta}G_{-\frac{1}{2}}H_{-\frac{1}{2}}|\phi\rangle\Big). \\  
\end{split}\ee
The second term in \refb{w} is the descendant of the null state at level $1/2$. So if we set $|1/2\rangle=0$, then we are left with $M_{-1}|\phi\rangle$ at level $1$. 

\bigskip

\paragraph {\it{Level 3/2:}} At this level, we have 6 constants $d_1,d_2, \ldots d_6$ satisfying equations: 
\bea
d_2\xi+d_5=0, \quad  d_2(1+\Delta)+d_4\xi+d_6 =0, \\ 
d_5\Delta+(4d_2+d_6)\xi+\frac{1}{3}\Big(c_Ld_5+c_Md_6\Big) =0, \quad d_5\Big(\xi+\frac{c_M}{3}\Big)=0,
\eea
and $d_1=0$; $d_2=d_3$. Considering the case where $c_M=0$, we find that to get a non-trivial state at this level, $\xi=d_1=d_5=0$. The null state is given by:
\be\label{x}
|3/2\rangle=d_2[L_{-1}H_{-\frac{1}{2}}|\phi\rangle+M_{-1}G_{-\frac{1}{2}}|\phi\rangle-(1+\Delta)H_{-\frac{3}{2}}|\phi\rangle] +d_4M_{-1}H_{-\frac{1}{2}}|\phi\rangle.
\ee
Except for the third term in \refb{x}, all the other terms are descendants of the null state at level $1/2$ and $1$. Setting $|1/2\rangle=|1\rangle=0$, we are left with $H_{-\frac{3}{2}}|\phi\rangle$ at level $3/2$. 

\bigskip

\paragraph{\it{Level 2:}}  If we set $c_M=0$, we have 9 constants $f_1,f_2, \ldots f_9$ which follow equations given in Appendix \ref{apd}. For $\Delta\neq0$ and $c_L\neq\frac{9}{2}$, the null state at this level is:
\bes
|2\rangle&=f_3L_{-1}M_{-1}|\phi\rangle 
+f_7\Big[M_{-1}G_{-\frac{1}{2}}H_{-\frac{1}{2}}|\phi\rangle-\frac{3+2\Delta}{2}M^2_{-1}|\phi\rangle\Big] \\
&+f_8\Big[G_{-\frac{3}{2}}H_{-\frac{1}{2}}|\phi\rangle-G_{-\frac{1}{2}}H_{-\frac{3}{2}}|\phi\rangle+\Big(2+\frac{4\Delta}{3}\Big)M_{-2}|\phi\rangle\Big]. \\
\end{split}\ee
The same analysis can be done here and we find that except for $M_{-2}|\phi\rangle$, all the remaining terms are descendants of the null states at lower levels. We can thus set $M_{-2}|\phi\rangle=0$ and carry on doing the same exercise for higher and higher levels. This means we can throw away the $H$'s at all half-integer levels and the $M$'s at integer levels. This truncates the algebra to $L$ and $G$ s, leaving us with a single copy of Super-Virasoro algebra.

\section{Concluding Remarks}

\subsection*{Summary}
In this paper, we have discussed a theory of parity violating $\mathcal{N}=2$ supergravity in asymptotically flat spacetimes and its dual field theory.  We looked at a scaling limit of this theory where the asymptotic symmetries from the bulk perspective reduce to a single copy of the Super-Virasoro algebra from the parent ``despotic" Super BMS algebra, borrowing nomenclature from \cite{Lodato:2016alv}, or the Inhomogeneous Super Galilean Conformal Algebra. Through a study of null vectors in the putative dual 2d field theory, we showed that this feature is also mirrored on the boundary. We call this the chiral reduction of the SGCA$_I$. The bulk theory correspondingly is called Flat Chiral Supergravity. The principal claim of this paper is thus the following new holographic connection: 
\paragraph{\it{Holographic correspondence:}} Flatspace Chiral Supergravity, defined by the action \refb{chiral_action} and boundary conditions \refb{bc}, is dual to a 2d chiral superconformal field theory with central charge $c=3 \kappa$.

\subsection*{SGCA$_H$ or the Democratic limit}
Interestingly, there exists another variant of the supersymmetric GCA, called the Homogeneous SGCA or SGCA$_H$, which arises from the analysis of asymptotic symmetries of supergravity on flat spacetimes \cite{Barnich:2014cwa} and also in the analysis of tensionless superstrings \cite{Lindstrom:1990ar, Bagchi:2016yyf}. Here the fermions are scaled in the same way \cite{Bagchi:2016yyf} and was called the ``democratic" limit in \cite{Lodato:2016alv}. This algebra is given by 
\bea\label{sgcah}
&& [L_n, L_m] = (n-m) L_{n+m} + \frac{c_L}{12} \, (n^3 -n) \delta_{n+m,0}, \nonumber\\
&& [L_n, M_m] = (n-m) M_{n+m} + \frac{c_M}{12} \, (n^3 -n) \delta_{n+m,0}, \\
&& [L_n, Q^\a_r] = \Big(\frac{n}{2} - r\Big) Q^\a_{n+r}, \quad \{Q^\a_r, Q^\b_s \} = \delta^{\a\b} \left[M_{r+s} + \frac{c_M}{6} \Big(r^2 - \frac{1}{4}\Big)  \delta_{r+s,0} \right]. \nonumber
\eea
In the analysis of \cite{Lodato:2016alv}, the central charge $c_L$ was zero as is expected from usual supergravity. But we can introduce a non-zero $c_L$ by using the same method as described earlier in this paper. 

\paragraph{\it Chiral truncation?} We could have asked whether the usual theory of supergravity in 3d flat spacetimes modified in aforementioned way, of which \refb{sgcah} is the asymptotic symmetry algebra, admits a chiral truncation as the one we have just seen, when we tune $c_M$ to zero. Here we notice that the above algebra \refb{sgcah} {\em{does not}} admit a super-Virasoro sub-algebra and so a truncation down to the chiral half of a superconformal theory is not possible. One could wonder whether there is a truncation down to just a Virasoro algebra. This stems from the observation that chiral truncation in the bosonic sector essentially amounted to setting the $M$'s to zero. From the above algebra it seems that since $\{Q, Q\}$ closes to $M$, setting the $M$'s to zero would also set all the supersymmetry to zero. This seems to be a rather unsatisfactory situation. A truncation in a supersymmetric theory leading to a theory without supersymmetry is unusual. This is especially true when one considers the case of tensionless superstrings, as considered in \cite{Bagchi:2017cte}. But an analysis of null states in this algebra carried out in \cite{Bagchi:2017cte} indicates that this truncation is not an allowed truncation. We are yet to understand what prevents this truncation from the point of view of the bulk and we leave this to future work. 

\paragraph{\it Emergent R-symmetry:} Towards a better understanding of the SGCA$_H$, we make a curious observation before finishing. If one switches $c_M =0$ but allows a finite $c_L$ in \refb{sgcah}, there is an emergent $U(1)_k$ R-symmetry admitted by the algebra. Let us denote the modes of this new current algebra by $P_n$. Then the non-trivial brackets of $P_n$ with the rest of the generators of \eqref{sgcah} are:
\begin{align}
		[L_n,P_m]=-mP_{n+m}, \, \, [P_n,P_m]=k\,n\delta_{n+m,0}, \,\, [P_n, Q^{\alpha}_r]=i\,\epsilon ^{\alpha \beta} Q^{\beta}_{n+r}.
\end{align}
This algebra also allows a 1-parameter spectral flow. This becomes manifest by the following basis change of the supercharges:
\begin{equation}
Q^{\pm} = \frac{1}{\sqrt{2}} \left( Q_1 \pm i\, Q_2\right)
\end{equation}
Note, now $Q^{\pm}_r$ have definite charges $\pm 1$ under $P_0$. And also now:
\begin{eqnarray} \label{anticom}
\{ Q^{+}_{r}, Q^{-}_{s}\} = M_{r+s} , \,\, \{ Q^{+}_{r}, Q^{+}_{s}\} = 0 =\{ Q^{-}_{r}, Q^{-}_{s}\}
\end{eqnarray}
Then the relabeling of the generators as:
\begin{eqnarray} \label{sp_flow}
&&\tilde{Q}^{\pm}_r = Q^{\pm}_{r \pm \eta}, \, \, \tilde{L}_n = L_n + \eta P_n + \frac{\eta^2 k}{2} \delta_{n,0} \non\\
&& \tilde{P}_n = P_n + \eta k \delta_{n,0} , \, \, \tilde{M}_n = M_n
\end{eqnarray}
turns out to be an inner-automorphism of \eqref{sgcah} with the $U(1)_k$ current, thus this leads to a spectral flow. It would be interesting to use this spectral flow symmetry of the super-BMS algebra \eqref{sp_flow} in a way analogous to \cite{Basu:2017aqn}, in holographic context.

\subsection*{Flatspace Chiral Supergravity with more SUSY?}
Finally, let us comment on a natural extension of the results of this paper. It is interesting to ask whether the $\mathcal{N}=2$ theory that we have just discussed is the only supersymmetric theory where we can observe such truncations in the bulk and boundary theories. We believe this is not the case. We can take, e.g. the $\mathcal{N}=4$ extended Super-BMS theory constructed in \cite{Basu:2017aqn}, the underlying symmetry algebra of which is given by: 
\begin{subequations}
\bea\label{n4}
&& [L_n, L_m] = (n-m) L_{n+m} + \frac{c_L}{12} \, (n^3 -n) \delta_{n+m,0}, \\
&& [L_n, M_m] = (n-m) M_{n+m} + \frac{c_M}{12} \, (n^3 -n) \delta_{n+m,0}, \\
&& [L_n, G^\pm_r] = \Big(\frac{n}{2} - r\Big) G^\pm_{n+r}, \ [L_n, R^\pm_r] = \Big(\frac{n}{2} - r\Big) R^\pm_{n+r}, \ [M_n, G^\pm_r] = \Big(\frac{n}{2} - r\Big) H^\pm_{n+r} \\
&& [L_n, J_m] = -m  J_{n+m}, \quad [L_n, P_m] = -m  P_{n+m}, \quad [M_n, J_m] = -m  P_{n+m}, \\
&& [J_n, G^\pm_r] = \pm G^\pm_{n+r}, \quad [J_n, R^\pm_r] = \pm R^\pm_{n+r}, \quad  [P_n, G^\pm_r] = \pm R^\pm_{n+r} \\
&& \{G^\pm_r, G^\mp_s \} = 2L_{r+s} \pm (r-s) J_{r+s} + \frac{c_L}{3} \Big(r^2 - \frac{1}{4}\Big)  \delta_{r+s,0} \\
&& \{G^\pm_r, R^\mp_s \} = 2M_{r+s} \pm (r-s) P_{r+s} + \frac{c_M}{3} \Big(r^2 - \frac{1}{4}\Big)  \delta_{r+s,0} \\
&& [J_n, J_m] = \frac{c_L}{3} n \delta_{n+m, 0}, \quad [J_n, P_m] = \frac{c_M}{3} n \delta_{n+m, 0}
\eea
\end{subequations}
The suppressed commutators, as usual, are zero. The initial indications are that we should be able to consistently ``turn off" $M_n, R_n^\pm,$ and  $P_n$ {\footnote{The first check of this is to see if putting these charges to zero leads to a consistent reduced algebra.}}. This would lead us to a chiral $\mathcal{N}=4$ Super-Virasoro algebra, generated by $L_n, G^\pm_n, J_n$. Of course, one needs to do the analogue of the analysis that we performed in the bulk and also the full null state analysis of this algebra to check whether this truncation is consistent. But the indications are that this would again work and should lead to a flatspace chiral supergravity, now with more supersymmetry.  
\bigskip

\subsection*{Acknowledgements}
Discussions with Glenn Barnich, Daniel Grumiller, and Max Riegler are gratefully acknowledged. We would like to acknowledge the warm hospitality of the following institutes and universities during various stages of this work: Universite Libre de Bruxelles (AB, PP), Albert Einstein Institute (PP), Vienna University of Technology (AB), University of Southampton (AB). The work of AB is partially supported by a DST Inspire award, a Max-Planck DST mobility award, and an India-Austria bilateral research project. RB acknowledges support by DST (India) Inspire award and in part the Belgian Federal Science Policy Office (BELSPO) through the Interuniversity Attraction Pole P7/37 and a research-and-return grant, in part by the “FWO-Vlaanderen” through the project G020714N, and by the Vrije Universiteit Brussel through the Strategic Research Program “High-Energy Physics”.
SD is a Research Associate of the Fonds de la Recherche Scientifique F.R.S.-FNRS (Belgium) and is supported in part by the ARC grant ``Holography, Gauge Theories and Quantum Gravity: Building models of quantum black holes''. 
He is also supported by IISN - Belgium (convention 4.4503.15) and benefited from the support of the Solvay Family.

\bigskip

\begin{appendix}
\section*{APPENDICES}
\section{Covariant phase space of Chern-Simons theory and global charges} \label{secpre-symp}
Most of the generic results presented in this subsection can be compared to \footnote{Some of the results, for example those stemming from the choice about the linear field dependence of the gauge parameter as presented in eq. (38) of \cite{Banados:1994tn}, are not generic enough to describe our purpose. This will be pointed out later.} \cite{Banados:1994tn} worked out in canonical formalism or can be extracted from more formal cohomological results in \cite{Barnich:1994db, Barnich:2000zw}. We present this section so as to make the paper self-consistent, in a formalism (for particular application of this formalism in the context of 3D bosonic gravity see \cite{Basu:2011qy}) which best suits are purpose of asymptotic charge calculation.

First variation of the Chern-Simons Lagrangian 3-form gives us the pre-symplectic potential $\Theta$, which is a two form on space-time and 1-form on covariant phase space \cite{Julia:2002df, Crnkovic:1986ex, Crnkovic:1987tz, Crnkovic:1986be, Ashtekar:1990gc} $\mathcal{P}$ (space of solutions):
\bea
\Theta (\delta) = -\frac{k}{4 \pi} \langle A \wedge \delta A \rangle .
\eea
Here, the variation $\delta$ serves as a vector field on $\mathcal{P}$ and in the above, it is contracted with $ \Theta$. It can be shown that the integrated (over some spatial 2-surface $\Sigma$ \footnote{In Chern-Simons theory framework, `spatial surface' does not hold much meaning as we don't require presence of any Lorentz structure or a metric. Only requiring that the background 3-manifold can be foliated as $T \times \Sigma$ for some real interval $T$ is sufficient (but not necessary).}) exterior derivative (with respect to $\mathcal{P}$) of $\Theta$ gives a background independent and closed pre-symplectic structure:
\bea
\Omega(\delta_1, \delta_2) = \frac{k}{4 \pi}\int_{\Sigma} \left( \delta_1 \Theta (\delta_2)- \delta_2 \Theta (\delta_1) \right).
\eea
In the above, the pre-sympletic 2-form is contracted with respect to two arbitrary variations or vector fields $\delta_1, \delta_2$.
For variations, which commute with each other, the expression simplifies to:
\bea \label{pre-symp}
\Omega(\delta_1, \delta_2) = \frac{k}{2 \pi}\int_{\Sigma} \langle \delta_1 A \wedge \delta_2 A \rangle
\eea

Now let us consider a particular diffeomorphism on $\mathcal{P}$ generated by the vector field $\delta_\Lambda$ on $\mathcal{P}$ such that it acts as gauge transformation of the connections:
\bea \label{gauge_trans}
\delta_\Lambda A = d \Lambda + [A, \Lambda],
\eea
($\lambda$ clearly is a Lie-algebra valued space-time function). For the present purpose we would consider only those gauge parameters $\Lambda$ which do not depend on field configurations (state-independent) in the bulk, but may do so in the boundary $\partial \Sigma$. Hence the form \eqref{pre-symp} of the pre-symplectic structure when supported over $\Sigma$ is justified.

Contracting $\delta_{\Lambda}$ with $\Omega$ gives:
\bea \label{cs_charge}
\Omega(\delta, \delta_{\Lambda}) = -\frac{k}{2 \pi}\int_{\partial \Sigma} \langle \Lambda, \delta A \rangle 
\eea
If $\Lambda$ continue to be state ($A$) independent even on $\partial \Sigma$, the above \footnote{In the explicit calculation, the space-time 1 form $\delta A$ is to be pulled back to the co-dimension 2 submanifold $\partial \Sigma$} expression is integrable (w.r.t $\delta$) trivially, to give the conserved charge:
\bea \label{integrable}
Q_0[\Lambda] = -\frac{k}{2 \pi}\int_{\partial \Sigma} \langle \Lambda, A \rangle
 \eea
modulo additive terms which are phase space constants. Here and always in this article we assume that if an integrated charge corresponding to some $\Lambda$ is found, it does not diverge at the boundary. This expression contains an integral supported only on $\partial \Sigma$; hence truly captures the fact that gauge transformations give rise to non-vanishing conserved charges through only boundary contributions. Extensions of $A$ in the bulk are redundant information. Hence the physical phase space $\tilde{\mathcal{P}} \subset \mathcal{P}$ of Chern-Simons theory, on which the charges \eqref{cs_charge} act, contain information of flat connections $A$ at the boundary. 

However, for the case of $\Lambda$ being state-dependent, right hand side of \eqref{cs_charge} is not an exact form on $\tilde{\mathcal{P}}$ and we would write that as an unintegrated $\slashed{\delta}Q$. 
To illustrate the point, let us consider for example $\Lambda$ as a linear function of $A$ which is the most simple non-trivial dependence:
\bea \label{lambda_def}
\Lambda = \Xi^{\mu}A_{\mu}  + \alpha
\eea
for some space-time vector field $\Xi$ which my have Bosonic as well as Fermionic components. 
 Note that $\Xi $ and the Lie-algebra valued space-time function $\alpha$ are both constants on $\tilde{\mathcal{P}}$. Interestingly the above gauge parameter $\Lambda$, on-shell induces diffeomorphism (and local supersymmetry transformation) by vector field $\Xi$ in addition to a state independent gauge transformation by $\alpha$. Now let us choose $\Sigma$ to be of the topology of a disc \footnote{The inclusion of black holes may be seen as inducing annular topology, which essentially modifies the homology of $\Sigma$, a canonical analysis of which can be found in \cite{Elitzur:1989nr}} and choose standard coordinates $r, \phi$ on it and $u$ (to be interpreted as retarded time coordinate in gravity setting) as the coordinate designating each $\Sigma$ foliation. Then the above unintegrated charge takes the form:
\bea \label{obstruc}
\slashed{\delta}Q[\Lambda] = -\frac{k}{2 \pi}\int_{\partial \Sigma = S^1} \, \langle \left(\Xi^u A_u + \Xi^r A_r \right), \delta A_{\phi} \rangle  + \delta \left(Q_0[\alpha] -\frac{k}{4 \pi}\int_{S^1} \Xi^{\phi} \langle A_{\phi}, A_{\phi}\rangle\right).
\eea
It is evident that the functions $A_u, A_r$ are obstructions against explicit integrability of the charge. As done in canonical analysis \cite{Banados:1994tn} in covariant phase space we cannot always do away with both $A_{u}$and $ A_{r}$ by gauge choice. This becomes clear by concentrating on the explicit example we worked with in the main body the article, particularly at \eqref{a}. As expected, the first part of \eqref{obstruc} is also the unintegrated form of charge associated with diffeomorphsim invariance. 
In case charges be integrated, we can at least formally calculate their Dirac brackets. That can be calculated directly from the pre-symplectic structure \eqref{pre-symp}
\bea \label{generic_db}
\Omega \left( \delta_{\Lambda_1}, \delta_{\Lambda_2} \right) = -\frac{k}{2 \pi} \int_{\partial_{\Sigma}} \left(\langle [\Lambda_1, \Lambda_2] , A\rangle + \langle \Lambda_2 , d \Lambda_1 \rangle \right)
\eea
This is the covariant phase space equivalent of the canonical Dirac bracket $\{ Q[\Lambda_1], Q[\Lambda_2]\}$. We must note an important caveat here that as we are not yet in a position to find expression of integrated charges, the above Dirac bracket calculation does not give us an explicit algebra of charges which might have been viewed  as dynamical realization of algebra of gauge transformation. 

As a side note let us take a closed look at the brackets of the charges due to state-independent gauge transformations $\alpha$. This can be readily extracted by putting $\Xi = 0$ in \eqref{lambda_def} and using \eqref{generic_db}. More specifically, define the charges:
\bea
J^I_m = Q_0[\alpha = e^{im \phi} T^I].
\eea
where $T$ are the Lie-algebra generators. Then the classical charge algebra becomes:
\bea
\{J^I_m , J^J_n \} = f^{IJ}{}_K J^K_{m+n}+ i k\, n \delta_{m+n,0} g^{IJ} 
\eea
the `classical' Kac-Moody algebra corresponding to the algebra on which the Chern-Simons theory was based. $f^{IJ}{}_K$  and $g_{IJ}$ respectively are the its structure constants and non-degenerate metric. 

\subsection{Reduced phase space} \label{red_ps}
Until now we have worked with the full gauge field $A$ and observed that its restriction at the asymptotic boundary is responsible for conserved global charges in the theory. Motivated by physically interesting scenarios \footnote{In the gravitational context these are boundary conditions coming from physically justified fall-off conditions of the connection fields}, which we particularly deal with in supergravity context of the present article, we will make further reduction of the phase space. The type of reduction, that would be useful for us is given by:
\bea \label{A-red}
A= b^{-1} \left( d + a\right)b.
\eea
In the coordinate chart that describes the background, we would choose $b = e^{r L}$ with $L$ being a fixed element in the Lie-algebra and $a$ is the pull-back of $bAb^{-1}$ to the surface $r=\mathrm{const.} \rightarrow \infty$.  
 Hence the functions $a^I_u, a^I_{\phi}$ ($I,J$ will denote Lie-algebra index in a chosen basis) span a reduced pace $\mathcal{P}_{\mathrm{red}} \subset \tilde{\mathcal{P}}$ defined by $\delta A_r =0.$ Let us now define $\lambda = b \Lambda b^{-1}$. Note that, since $a$ is the pull back of the 1-form $bAb^{-1}$ to the boundary, it is easy to see that $\lambda$ can be taken to be independent of $r$ and hence suitable to be a function defined exclusively on the boundary. Further, only with this choice of $\lambda$ the reduced phase space $ \mathcal{P}_{\mathrm{red}}$ is preserved. 

Now, for field dependent gauge transformations \eqref{lambda_def} this means:
\bea \label{small_lambda}
\lambda &=& \Xi^{\mu } \left( \partial_{\mu}b \, b^{-1} + a_{\mu} \right) + \beta ~~\mbox{where } \beta = b \alpha b^{-1} \non \\
&=& \Xi^u a_u + \Xi^\phi a_{\phi } + \Xi^r L + \beta
\eea
It now becomes clear from \eqref{obstruc} that on $\mathcal{P}_{\mathrm{red}}$, that the obstruction against integrability of corresponding charges comes as:
\bea \label{obstruc2}
\slashed{\delta}Q[\Lambda] &=& -\frac{k}{2 \pi}\int_{S^1} d\phi \langle \Lambda, \delta A_{\phi} \rangle = -\frac{k}{2 \pi}\int_{S^1} d\phi \langle \lambda, \delta a_{\phi} \rangle \non \\
&=& \mbox{integrable part } -\frac{k}{2 \pi}\int_{S^1} d\phi \,\Xi^{u }\langle a_u , \delta a_{\phi} \rangle
\eea
Basically the sufficient requirement for integrability of the charge, for any arbitrary vector fields $\Xi$ is integrability of the inner product $\langle a_u , \delta a_{\phi} \rangle$.

The up-shot of the present discussion is any diffeomorphism including local supersymmetry (or any linearly state dependent gauge transformation) gives rise to integrated conserved charge on $\mathcal{P}_{\mathrm{red}}$, provided the above sufficiency \footnote{If $y^i$ are the coordinates of the reduced phase space, which should be even in number and may well be space-time functions, then the integrability condition boils down to:
$$g_{IJ} \left(\frac{\delta a^I_u}{\delta y^i}\frac{\delta a^J_\phi}{\delta y^j} - \frac{\delta a^I_u}{\delta y^j}\frac{\delta a^J_\phi}{\delta y^i}\right) = 0 $$ for all $i,j$ and $g_{IJ}$ is the Lie-algebra metric} is met and the charge integrals don't diverge as $r \rightarrow \infty$. Moreover it also means that only the $a^I_\phi$ do span $\mathcal{P}_{\mathrm{red}}$ as $a^I_u$ must be functions of $a^I_\phi$ for the sufficiency. This is illustrated explicitly in the supergravity context as in \eqref{a}. 
%
Up to this point we have stated existence conditions of integrated charge on the reduced phase space $\mathcal{P}_{\mathrm{red}}$ for arbitrary $\Lambda(\Xi, \alpha)$. One can now use \eqref{generic_db} to compute Dirac brackets of these charges. However that might not lead to closed algebra of charges on $\mathcal{P}_{\mathrm{red}}$. In order to ensure that, we will have to consider only those transformations $\delta_{\Lambda}$ which are tangential vector fields on $\mathcal{P}_{\mathrm{red}}$, ie. preserves the condition \eqref{A-red} via $\partial_r \lambda = 0$ and preserves the integrability criterion of $\langle a_u , \delta a_\phi \rangle$. In the following explicit example we will reduce the phase space further employing physical boundary conditions. More restrictive choice of the gauge parameters should be taken in order to preserve them.
\subsection{Charge algebra on reduced phase space} \label{charge_algebra_proof}
We have outlined above that arbitrary diffeomorphisms can result into conserved boundary charges. However for a closed algebra of charges more stringent conditions come in to play. Before going to the explicit example of the starting supergravity theory, let us analyze the phase space in a bit more detail. Without losing generality, in what follows we would set $\alpha = 0 =\beta$ for \eqref{lambda_def}, \eqref{small_lambda}. Moreover we won't, at this stage put any additional boundary conditions which may reduce $\mathcal{P}_{\mathrm{red}}$ further. This means that all the components $a^I_{\phi}$ which we will denote as $y^I$ (which are function of $u, \phi$, the boundary coordinates) now on, do span $\mathcal{P}_{\mathrm{red}}$. For integrability we put an linear ansatz for the functional form \footnote{This can be contrasted with the situation discussed in \cite{Grumiller:2017sjh}, where all components of $ a_u$ are taken to be phase space constants and thereby ensuring integrability trivially} of $a_u (y)$ on $\mathcal{P}_{\mathrm{red}}$:
\bea \label{criterion}
a_u^I = C^I{}_J y^J 
\eea 
for both phase space and space-time constant automorphisms $C^I{}_J$ on the Lie-algebra. It should be kept in mind that the equations of motion, ie the flatness of the original connection $A$ is always implied. On $\mathcal{P}_{\mathrm{red}}$ this now gives:
\bea \label{red_eom}
C^I{}_J \p_\phi y^J - \p_u y^I + C^K{}_L y^J y^L f^I{}_{JK} = 0.
\eea
It is now easy to see that the expression:
\bea
\langle a_u , \delta a_\phi \rangle = \delta\left(\frac{1}{2}C_{IJ} y^I y^J \right)
\eea is integrable if and only if $C_{IJ}$ is a symmetric tensor on the Lie-algebra. The corresponding charge is now:
\bea
Q[\Xi] = -\frac{k}{2 \pi}\int_{\partial \Sigma } d\phi \left[\frac{1}{2} \mg^{(\Xi)}_{IJ}y^I y^J +  \Xi^r L^I y^J g_{IJ} \right]
\eea
where we have introduced:
$$\mg^{(\Xi)}_{IJ} = C_{IJ} \Xi^u + g_{IJ} \Xi ^{\phi}$$
with $g_{IJ}$ the Lie-algebra metric.
We should now consider the symmetries, which are allowed, in a sense of preserving $\mathcal{P}_{\mathrm{red}}$. This means also preserving the integrable structure \eqref{criterion} of charges on $\mathcal{P}_{\mathrm{red}}$, ie:
\bea \label{key}
&&\delta a_u^I = C^I{}_J \delta y^J \non \\
\mbox{hence }&& C^I{}_J \left( \p_{\phi} \lambda^J + y^K \lambda^L f_{KL}{}^J \right) = \p_{u} \lambda^I + C^K{}_J y^J \lambda^L f_{KL}{}^I .
\eea
It should be noted that this is they key equation for `allowed' set of gauge transformation. On the other hand this equation governs the physical boundary diffeomorphisms that preserve $\mathcal{P}_{\mathrm{red}}$, or asymptotic boundary conditions. This essentially is a subset of all possible boundary diffeomorphisms. We would also stress here that more strict set of boundary conditions can only reduce the allowed space of diffeomorphisms.
Using \eqref{key} and putting in the pre-symplectic form \eqref{generic_db}, we can now compute the Dirac bracket of charges corresponding to 2-gauge parameters $\lambda_a = \mg^{(\Xi_a)J}{}_I y^I T_J + \Xi_a L, ~ a=1,2 $:
\bea \label{final_alg}
\{ Q[\Xi_1], Q[\Xi_2] \} &=& -\frac{k}{2 \pi} \int_{\partial_{\Sigma}} \left(\langle [\Lambda_1, \Lambda_2] , A\rangle + \langle \Lambda_2 , d \Lambda_1 \rangle \right)\non \\&=&-\frac{k}{2 \pi} \int_{\partial \Sigma} \langle \lambda_2 ,  \left(\partial_{\phi} \lambda_1 + [a_{\phi} , \lambda_1] \right)\rangle \non \\
&=& Q[\tilde{\Xi}] + \frac{k}{2\pi } \int_{\partial \Sigma}  \Xi_1^r L^I \p_\phi (\Xi_2^r L_I )
\eea
where $\tilde{\Xi} = -[\Xi_1, \Xi_2]$ is the Lie bracket of the vector fields $\Xi_1, \Xi_2$ restricted on the $r=$ constant boundary surface. The steps involved in this calculation are a bit too lengthy to incorporate in the main body of the paper, and is given in the Appendix. The equation \eqref{final_alg} represents the dynamical realization of the algebra of allowed diffeomorphisms that preserve the reduced boundary phase space. The dynamical realization is not exact due to the presence of the central term $\int_{\partial \Sigma}  \Xi_1^r L^I \p_\phi (\Xi_2^r L_I )$
\section{Deriving \eqref{final_alg}}
We here supply the steps involved in deriving \eqref{final_alg}. We start with the criterion \eqref{criterion} every allowed gauge parameter or diffeomorphism must satisfy in order to preserve the reduced phase space.
 
While this is they key equation for `allowed' set of gauge transformation, more usable sets of information can be derived from it:
\bea \label{useful}
\p_u \mg^{(\Xi)}_{IJ}  &=& C_{IM} \left(\p_{\phi} \mg^{(\Xi)M}{}_J + y^K \mg^{(\Xi)L}{}_{(J} f_{K)L}{}^M \right) - C^K{}_{(J} \mg^{(\Xi)L}{}_{M)}y^M f_{KL}{}^I \non \\
				&+& \Xi^r L^L C_{M(I}  f_{J)L}{}^M \\
g_{IJ}\p_u \Xi^r &=& C_{IJ} \p_{\phi} \Xi^r +y^M \Xi^r (C_{IK} f_{MJ}{}^K+  C_{MK} f_{IJ}{}^K)				
\eea 
Here the symmetrization brackets are used without any combinatoric factor.
Let us now consider the following expression for the Lie bracket $\tilde{\Xi} = -[\Xi_1, \Xi_2]$:
\bea 
&&-\frac{1}{2}\int_{\partial \Sigma} \mg^{(\tilde{\Xi})}_{IJ} y^I y^J 
= \frac{1}{2}\int_{\partial \Sigma} \left[(\Xi_1^u \p_u + \Xi_1^\phi \p_\phi )\mg^{(\Xi_2)}_{IJ} - (\Xi_1 \leftrightarrow \Xi_2) \right]y^I y^J \non \\
&& = \int_{\partial \Sigma}  y^I\mg^{(\Xi_1)}_{IK} \p_\phi \left(\mg^{(\Xi_2)K}{}_{J}y^J\right) - \int_{\partial \Sigma}\left(\Xi_1^u \Xi_2^r- \Xi_1^r \Xi_2^u \right) L^L C_{IM}  f_{JL}{}^M  y^I y^J \non
\eea
Otherwise written this means:
\bea \label{term1}
\int_{\partial \Sigma}  y^I\mg^{(\Xi_1)}_{IK} \p_\phi \left(\mg^{(\Xi_2)K}{}_{J}y^J\right) = -\frac{1}{2}\int_{\partial \Sigma} \mg^{(\tilde{\Xi})}_{IJ} y^I y^J +\int_{\partial \Sigma}\left(\Xi_1^u \Xi_2^r- \Xi_1^r \Xi_2^u \right) L^L C_{IM}  f_{JL}{}^M  y^I y^J
\eea

On the other hand the following expression simplifies to:
\bea \label{term2}
-\int_{\partial \Sigma} \tilde{\Xi}^r L^I y^J g_{IJ}  &=& \int_{\partial \Sigma} y^I L^J \big[ \mg^{\Xi_1}_{IJ} \p_\phi \Xi_2^r + \Xi_1^u \Xi_2 ^r y^M (C_{IK} f_{MJ}{}^K+  C_{MK} f_{IJ}{}^K)  \non \\
&& + g_{IJ}\Xi_1^\phi \p_\phi\Xi_2 ^r \big] - (\Xi_1 \leftrightarrow \Xi_2) \non \\
&=&  \int_{\partial \Sigma} y^I \mg^{\Xi_1}_{IJ} \p_\phi (\Xi_2^r L^J) + \Xi_1^r L^I \p_\phi \left(  \mg^{\Xi_2}_{IJ} y^J\right)
\eea
Now from the expression \eqref{final_alg} the non-derivative term simplifies as:
\bea
-\int_{\partial \Sigma} \langle \lambda_2 , [a_{\phi} , \lambda_1]\rangle = \int_{\partial \Sigma}(\Xi_1^u \Xi_2^r - \Xi_1^r \Xi_2^u) L^L C_{IM } f_{JL}{}^M y^I y^J 
\eea
where we have used $\lambda^I_1 = \mg^{(\Xi_1)I}{}_J y^J + \Xi_1^r L^I$ and $\lambda^I_2 = \mg^{(\Xi_2)I}{}_J y^J + \Xi_2^r L^I$.

Hence the expression for the bracket of the charges \eqref{final_alg} can be written upto constant multipliers:
\bea
&&-\int_{\partial \Sigma} \langle \lambda_2 ,  \left(\partial_{\phi} \lambda_1 + [a_{\phi} , \lambda_1] \right)\rangle \non \\
&=& \int_{\partial \Sigma} y^I\mg^{(\Xi_1)}_{IK} \p_\phi \left(\mg^{(\Xi_2)K}{}_{J}y^J\right) + y^I \mg^{\Xi_1}_{IJ} \p_\phi (\Xi_2^r L^J) + \Xi_1^r L^I \p_\phi \left(  \mg^{\Xi_2}_{IJ} y^J\right) \non \\
&& + \Xi_1 ^r L^I \p_\phi (\Xi_2 ^r L_I ) + (\Xi_1^u \Xi_2^r - \Xi_1^r \Xi_2^u) L^L C_{IM } f_{JL}{}^M y^I y^J 
\eea 
It is now trivial to use the expressions in \eqref{term1}, \eqref{term2} to see that:
\bea
\int_{\partial \Sigma} \langle \lambda_2 ,  \left(\partial_{\phi} \lambda_1 + [a_{\phi} , \lambda_1] \right)\rangle = \int_{\partial \Sigma} \frac{1}{2}\mg^{(\tilde{\Xi})}_{IJ} y^I y^J +  \tilde{\Xi}^r L^I y^J g_{IJ} - \Xi_1 ^r L^I \p_\phi (\Xi_2 ^r L_I )
\eea
Hence the bracket of charges corresponding to the vector fields $\Xi_1$ and $\Xi_2$ is
\bea
\{ Q[\Xi_1], Q[\Xi_2] \} &=& -\frac{k}{2\pi} \int_{\partial \Sigma} \langle \lambda_2 ,  \left(\partial_{\phi} \lambda_1 + [a_{\phi} , \lambda_1] \right)\rangle \non\\
&=& Q[\tilde{\Xi}= - [\Xi_1, \Xi_2]] + \frac{k}{2\pi } \int_{\partial \Sigma}  \Xi_1 ^r L^I \p_\phi (\Xi_2 ^r L_I )
\eea

\section{Explicit computation of the Dirac bracket for Supergravity}
We would present the explicit computation \eqref{sample} here. According to the the definitions of these modes \eqref{modes}, we see that the charges $L_m$ correspond to the gauge transformation whose asymptotic form is given by $\lambda(Y_1=e^{im \phi},0,0,0)$. 

The Dirac bracket follows from the formula derived in \eqref{generic_db}. For this we should be considering two gauge transformations $\lambda_1$ and $\lambda_2$ such that among their component functions, only $\chi^{+1}(\phi) = Y(\phi)$ do survive. $Y_1=e^{im \phi}, Y_2=e^{in \phi} $ give the specific modes given above. Using \eqref{lambda_components}, \eqref{lambda_components_more}, we get:
\begin{align} \label{lambda_form}
\lambda_1 &= Y_1 L_1 - Y'_1 L_0 + \left(\frac{1}{2}Y''_1 - \frac{\mathcal{M}}{4}Y_1 \right)L_{-1} + u Y'_1 M_1 - u Y''_1 M_0 \non \\
& +\left(\frac{1}{2} u Y'''_1 -\frac{\mathcal{M}}{4}u Y'_1 -\frac{\mathcal{N}}{4} Y_1   \right) M_{-1} + \frac{1}{4} \left( Y_1 \psi \, G_{-1/2}+ \left( Y_1 \eta + uY'_1 \psi\right) R_{-1/2}\right)
\end{align}
and similar for $\lambda_2$. For the expression in \eqref{generic_db}, we first evaluate the first term using the brackets \eqref{alg}, \eqref{alg2} of the algebra $\tilde{\mathfrak{g}}$ and the inner product \eqref{inn_prod}, \eqref{mod_inn_prod}. A few lines of algebraic manipulation yields 
\bea \label{checksum} \int_{S^1}\langle [\lambda_1, \lambda_2], a\rangle = \frac{2}{\mu} \int_{S^1}Y'_1 Y''_2
\eea
Note that, this term is independent of any phase space variable.

On the other hand the second term evaluates to:
\bea
\int_{S^1}\langle \lambda_1, d \lambda_2 \rangle = \frac{1}{2}  \int_{S^1} d \phi (-Y'_1 Y_2 + Y_1 Y'_2) \left(\mathcal{J} + \frac{1}{\mu} \mathcal{M} \right) + \frac{3}{\mu} \int_{S^1}Y'_1 Y''_2
\eea
Hence, 
\bea
\{Q[\lambda_1(Y_1,0,0,0)] , Q[\lambda_2(Y_2,0,0,0)] \} &=& -\frac{k}{2 \pi} \int_{\partial_{\Sigma}} \left(\langle [\Lambda_1, \Lambda_2] , A\rangle - \langle \Lambda_1 , d \Lambda_2 \rangle \right)\non \\
&=&-\frac{k}{2 \pi} \int_{\partial_{\Sigma}} \left(\langle [\lambda_1, \lambda_2] , a\rangle - \langle \lambda_1 , d \lambda_2 \rangle \right)\non \\
&=& -\frac{k}{4 \pi}\int_{S^1} d \phi (Y'_1 Y_2 - Y_1 Y'_2) \left(\mathcal{J} + \frac{1}{\mu} \mathcal{M} \right) \non \\ &&+ \frac{k}{2 \pi} \frac{1}{\mu} \int_{S^1}Y'_1 Y''_2
\eea
The Dirac bracket for the modes are found by setting $Y_1 = e^{im \phi}, Y_2 = e^{in \phi}$.

\section{Details of the Null state analysis} \label{apd}
This appendix contains some of the detailed calculations of Sec 3, which we omitted in the main text. First we list the conditions of the states at level $\frac{3}{2}$ and level 2 being null.  \\
Level 3/2:
\begin{equation}
\begin{split}
G_{\frac{1}{2}}|3/2\rangle&=[d_1(1+2 \Delta)+2d_2\xi+2d_5]L_{-1}|\phi\rangle+2[d_3(1+ \Delta)+d_4\xi+d_6]M_{-1}|\phi\rangle \\
&+(d_2-d_3)G_{-\frac{1}{2}}H_{-\frac{1}{2}}|\phi\rangle=0 \\
H_{\frac{1}{2}}|3/2\rangle&=2d_1\xi L_{-1}|\phi\rangle+2[d_1+d_3\xi+d_5]M_{-1}|\phi\rangle \\
&+2d_1G_{-\frac{1}{2}}H_{-\frac{1}{2}}|\phi\rangle=0 \\
L_1|3/2\rangle&=[d_1(1+2 \Delta)+2d_3\xi+2d_5]G_{-\frac{1}{2}}|\phi\rangle \\
&+[d_2(1+2 \Delta)+d_3+2d_4\xi+2d_6]H_{-\frac{1}{2}}|\phi\rangle=0 \\
M_1|3/2\rangle&=d_1(1+2\xi)G_{-\frac{1}{2}}|\phi\rangle+2(d_2\xi+d_5)H_{-\frac{1}{2}}|\phi\rangle=0 \\
G_{\frac{3}{2}}|3/2\rangle&=[2(2d_1+d_5) \Delta+2(2d_2+2d_3+d_6)\xi+\frac{2}{3}(c_Ld_5+c_Md_6)]|\phi\rangle=0 \\  
H_{\frac{3}{2}}|3/2\rangle&=[2(2d_1+d_5)\xi+\frac{2d_5c_M}{3}]|\phi\rangle=0 \\  
\end{split}
\end{equation}
Level 2:
\begin{equation}
\begin{split}
G_{\frac{1}{2}}|2\rangle&=2(f_2-f_4\xi)L_{-1}G_{-\frac{1}{2}}|\phi\rangle+[f_3+2f_4(1+ \Delta)+2f_8]L_{-1}H_{-\frac{1}{2}}|\phi\rangle
+(f_3-2f_7\xi-2f_9)M_{-1}G_{-\frac{1}{2}}|\phi\rangle \\ &+[2f_5+f_7(3+2 \Delta)]M_{-1}H_{-\frac{1}{2}}|\phi\rangle +\Big(\frac{3f_1}{2}-2f_8\xi \Big)G_{-\frac{3}{2}}|\phi\rangle+\Big[\frac{3f_6}{2}+f_9(3+2 \Delta)\Big]H_{-\frac{3}{2}}|\phi\rangle=0 \\  
H_{\frac{1}{2}}|2\rangle&=2(f_2+f_4\xi)L_{-1}H_{-\frac{1}{2}}|\phi\rangle+[f_3+2(f_4+f_7\xi+f_8)]M_{-1}H_{-\frac{1}{2}}|\phi\rangle +\Big(\frac{3f_1}{2}+2f_9\xi\Big)H_{-\frac{3}{2}}|\phi\rangle=0 \\  
L_1|2\rangle&=[3f_1+2f_2(1+2 \Delta)+2(f_3+f_4)\xi]L_{-1}|\phi\rangle +[2f_3(1+ \Delta)+2(f_5+f_7)\xi+3f_6+2f_9]M_{-1}|\phi\rangle \\
&+[2f_4(1+ \Delta)+2f_7\xi+2f_8+2f_9]G_{-\frac{1}{2}}H_{-\frac{1}{2}}|\phi\rangle=0 \\  
M_1|2\rangle&=4f_2\xi L_{-1}|\phi\rangle+(3f_1+2f_2+2f_3\xi)M_{-1}|\phi\rangle+2f_4\xi G_{-\frac{1}{2}}H_{-\frac{1}{2}}|\phi\rangle=0 \\  
G_{\frac{3}{2}}|2\rangle&=\Big[\frac{5f_1}{2}+2f_2+4f_4\xi+2f_9\Big(\xi+\frac{c_M}{3}\Big)\Big]G_{-\frac{1}{2}}|\phi\rangle \\
&+2\Big[f_3+f_4(1+2 \Delta)+\frac{5f_6}{4}+2f_7\xi+f_8\Big(\Delta+\frac{c_L}{3}\Big)+2f_9\Big]H_{-\frac{1}{2}}|\phi\rangle=0 \\  
H_{\frac{3}{2}}|2\rangle&=\Big[\frac{5f_1}{2}+2f_2+4f_4\xi+2f_8(\xi+\frac{c_M}{3})\Big]H_{-\frac{1}{2}}|\phi\rangle=0 \\  
L_2|2\rangle&=\Big[2(2f_1+3f_2) \Delta+(6f_3+6f_4+4f_6+6f_7+5f_8+3f_9)\xi+c_M\Big(f_9+\frac{f_6}{2}\Big)\Big]|\phi\rangle=0 \\  
M_2|2\rangle&=\Big[f_1\Big(4\xi+\frac{c_M}{2}\Big)+6f_2\xi\Big]|\phi\rangle=0 \\
\end{split}
\end{equation}

\paragraph{Null state conditions for level 2:} We now give the details of the null state conditions for level 2. If we set $c_M=0$, 
we have 9 constants $f_1,f_2...f_9$ satisfying equations:
\bes
f_3\xi=f_8\xi=f_9\xi&=0 \\  
f_3+2f_4(1+\Delta)+2f_8&=0 \\
f_3-2f_7\xi-2f_9&=0 \\
2f_5+f_7(3+2\Delta)&=0 \\
\frac{3f_6}{2}+f_9(3+2\Delta)&=0 \\  
f_3+2(f_4+f_7\xi+f_8)&=0 \\
2f_3(1+\Delta)+2(f_5+f_7)\xi+3f_6+2f_9&=0 \\
2f_4(1+\Delta)+2f_7\xi+2f_8+2f_9&=0 \\  
f_3+f_4(1+2\Delta)+\frac{5f_6}{4}+2f_7\xi+f_8\Big(\Delta+\frac{c_L}{3}\Big)+2f_9&=0 \\
(6f_3+4f_6+6f_7+5f_8+3f_9)\xi&=0 \\  
\end{split}\ee
and $f_1=f_2=0$. 
Demanding a non-trivial solution, we get $\xi=f_1=f_2=0$, $f_3=2f_9$. For the rest of the analysis, the reader is redirected to the main text.

\end{appendix}

\end{document}